\documentstyle[multicol,epsfig,prl,aps]{revtex}
\tighten
\draft

\begin{document}
\title{Transition from overscreening to underscreening in the 
multichannel Kondo model: exact solution at large N}  
 
\author{Olivier Parcollet, Antoine Georges} 
\address{ Laboratoire de Physique Th{\'e}orique de l'Ecole 
Normale Sup{\'e}rieure,\cite{auth1}
24 rue Lhomond, 75231 Paris Cedex 05, France}
\date{\today}
\maketitle
\begin{abstract}
A novel large-N limit of the multichannel Kondo model is introduced, 
for representations of the impurity spin described by   
Schwinger bosons. Three cases are found, associated with 
{\it underscreening}, {\it overscreening} and 
{\it exact Kondo screening} of the impurity. 
The saddle-point equations derived in this limit are reminiscent of 
the ``non-crossing approximation'', but preserve the Fermi-liquid 
nature of the model in the exactly screened case.   
Several physical quantities are computed, both numerically, and analytically 
in the low-$\omega,T$ limit, and compared to other approaches.   
\end{abstract}
\pacs{Pacs numbers: 75.20 Hr, 75.30 Mb, 71.10.Hf}

\def\Im{\mathop{\hbox{Im}}}
\def\ES{(ES)\,} 
\def\OS{(OS)\,} 
\def\US{(US)\,}

\begin{multicols}{2}

Besides their experimental relevance to magnetic impurities
in metals, heavy fermion compounds, and tunneling in
mesoscopic systems \cite{Moriond}, quantum impurity models provide
useful testing grounds for theoretical methods dealing with 
correlated electron systems. Among those, explicit solutions in the 
limit of large spin degeneracy ({\it e.g} SU($N$) spins in the large N
limit) have often proven to retain many crucial aspects of the     
low-energy physics while being simple to implement \cite{largen}.

In this letter, we introduce a novel large-N approach to the multichannel 
Kondo model \cite{NB,CZ}.
The non-Fermi liquid properties of this model in the overscreened 
regime have attracted considerable attention recently.  
Our approach is related to the earlier work of 
Cox and Ruckenstein \cite{CR}, in that we also consider the SU($N$) Kondo
model with $K$ channels of conduction electrons and take the limit 
$N,K\rightarrow\infty$ with $K/N=\gamma$ being fixed.    
There are two crucial differences with Ref.\cite{CR} however:   
{\it i)} we are dealing here with {\it Schwinger boson} representations of 
the impurity spin, in contrast with the fermionic representations 
considered in \cite{CR} and {\it ii)} we keep track of the 
quantum number associated with the 
``size'' of the impurity spin by imposing a local 
constraint on the Schwinger bosons which is also taken to scale 
as $N$. Because of these two new features, our approach 
captures all three possible regimes of the multichannel Kondo model 
(and transitions between them):
the {\it underscreened} regime 
(in which the Kondo effect only partially screens 
the impurity spin), the {\it overscreened} regime (in which a non-Fermi 
liquid state is formed at low-energy), and an 
{\it exactly screened} Fermi-liquid regime in between. 

In the large-N limit, our model is solved by a set of two coupled 
integral equations. These equations resemble in structure 
the ``Non-Crossing Approximation'' (NCA) \cite{NCA,CR}, 
with some crucial differences however associated with 
points {\it i)} and {\it ii)} above. Because of our handling 
{\it ii)} of the constraint, these equations result from a     
true saddle-point principle, with controllable fluctuations 
in $1/N$. Furthermore, in the exactly screened case, 
the equations derived in this paper {\it preserve the local Fermi-liquid 
character of the problem} down to zero-temperature and frequency.
For this reason, and because Schwinger boson mean-field theories 
yield  a quite satisfactory
description of magnetically ordered phases \cite{AA}, 
our approach offers new prospects for a 
successful treatment of the Kondo lattice model. 
 
We consider a generalised Kondo model with $K$ channels of conduction 
electrons and a spin symmetry group extended from $\mbox{SU}(2)$ to
$\mbox{SU}(N)$. An impurity spin $\vec{S}$ is placed at the origin. 
We choose to represent the $N^2-1$ components 
$(S_{\alpha\beta})_{1\leq\alpha,\beta\leq N}$ of $\vec{S}$ 
in terms of $N$ Schwinger bosons $b_{\alpha}$ with a constraint, namely:
\begin{equation}
S_{\alpha\beta} = b^\dagger_{\alpha}b_{\beta}-{P\over N}\,\delta_{\alpha\beta}
\,\,\,,\,\,\,
\sum_{\alpha=1}^{N} b^\dagger_{\alpha}b_{\alpha} \,=\, P
\label{spinboson}
\end{equation}
In all the following, the conduction electrons transform under the 
fundamental representation of $SU(N)$. 
The hamiltonian of the model reads:
\begin{equation}
H\, =\, \sum_{\vec{p}}\sum_{1\leq i \leq K \atop 1 \leq \alpha \leq N}
\epsilon_{\vec{p}} c^\dagger_{\vec{p}i\alpha} c_{\vec{p}i\alpha}
\,+\,J_{K} \sum_{\vec{p}\vec{p'}i\alpha\beta}
S_{\alpha\beta}\, \, { c^\dagger_{\vec{p}i\beta} c_{\vec{p'}i\alpha}}
\label{ham}
\end{equation}
In the usual $N=2$ case ($\alpha,\beta=\uparrow,\downarrow$), 
this is the multichannel Kondo model with 
an impurity spin of size $S=P/2$. For arbitrary $N$, Eq.(\ref{spinboson}) 
means that we have restricted ourselves to representations of $SU(N)$ 
corresponding to a Young tableau with a single line of $P$ boxes. 
Quantum fluctuations are stronger at small values of $P$, while large 
$P$ (for fixed $N$) corresponds to a semiclassical limit. 

It is easily checked that a weak antiferromagnetic coupling ($J_K>0$)
grows under renormalisation for all $K$ and $N$, and all representations
$P$ of the local spin. In order to
determine whether the renormalization group (RG) 
flow takes $J_K$ all the way to strong coupling
(underscreened or exactly screened cases), or whether an intermediate
non-Fermi liquid fixed point exists (overscreened case), we have generalised 
the Nozi{\`e}res and Blandin stability analysis of the strong-coupling 
fixed point $J_K=+\infty$ \cite{NB}. 
In this limit, a bound-state is formed between the impurity spin 
and the conduction electrons, which   
corresponds to a new spin representation 
dictated by the minimisation of the Kondo energy 
(second term in (\ref{ham})). One must then study the stability of 
this strong-coupling state when a small hopping $t$ of the conduction electrons 
is turned on. The technical steps involved in this analysis will be 
reported elsewhere \cite{long1,long2}, 
and only the main conclusions are summarized here. 
We have established that, for arbitrary value of  $N$, 
there are three possible strong-coupling regimes \cite{foot} depending 
on the number of channels $K$ as compared to $P$ :

{$\bullet$}
 When $P > K$, $(N-1)K$ conduction electrons bind to the impurity 
spin in such a way that a free local spin of size $P^{sc}=P-K$ is left 
unscreened for $J_K/t=\infty$. Turning on a hopping tends to increase the 
total spin of the system, corresponding to a weak {\it ferromagnetic} residual  
Kondo interaction as one departs from the strong-coupling fixed point. 
Hence, the latter is stable against this perturbation, and we have the 
typical situation of an {\it underscreened} \US Kondo effect.

{$\bullet$}
When $P=K$, $(N-1)K$ conduction electrons exactly screen the 
impurity spin and produce a spin singlet state ($P^{sc}=0$). The 
strong-coupling state, having the lowest possible degeneracy,   
must be again stable against $t/J_K$ and the model  
displays exact Kondo screening \ES .

{$\bullet$}
 When $P < K$, $(N-1)P$ conduction electrons  screen 
the impurity spin, while 
$(N-1)(K-P)$ arrange themselves such that a residual spin $P^{sc}=K-P$ 
remains. In contrast to the above case, the residual Kondo interaction 
produced when a hopping is turned on is now {\it antiferromagnetic}, and 
the strong-coupling fixed point is unstable. One expects, as confirmed 
below, that an intermediate fixed point exists with non-Fermi liquid 
properties: this is the {\it overscreened} \OS regime.

We now turn to the analysis of this model in the large-N limit. 
We shall proceed in a manner similar to Ref. \cite{CR},  
by setting $K=\gamma\,N$ and $J_K=J/N$ and taking the limit 
$N\rightarrow\infty$ for fixed values of
$\gamma$ and $J$. 
The crucial difference (apart from the use of Schwinger 
bosons) is that we shall deal with the constraint in 
Eq.(\ref{spinboson}) by setting $P=p_0\,N$ and keeping $p_0$ fixed  
(instead of fixing $P=1$ as in \cite{CR}). By doing so, we are preserving  
the existence of the transition in the large N limit, the three regimes 
above corresponding to $p_0>\gamma$ \US, $p_0=\gamma$ \ES and $p_0<\gamma$ 
\OS. This also insures that the model is controlled by a {\it true 
saddle-point} at large-N, with controllable $1/N$ corrections. 

In order to derive the saddle-point equations, we 
use a functional integral formulation of model (\ref{ham}),   
with a Lagrange multiplier field $\lambda(\tau)$ to implement the 
constraint. Conduction electrons can be  
integrated out in the bulk, keeping only degrees of freedom at the impurity
site. The local Kondo interaction is decoupled by introducing 
an auxiliary field in each channel $F_i(\tau)$, conjugate to the amplitude 
$\sum_{\alpha}c^\dagger_{i\alpha}(\tau)b_{\alpha}(\tau)$. This field will 
be responsible for capturing the physics of the Kondo effect. Note that it 
is a {\it Grassmanian (anticommuting) field}, because of our bosonic treatment 
of the impurity spin. 
After these manipulations, we are left with the effective action:
\begin{eqnarray}
\nonumber S\,&&=\,
\int^{\beta}_{0} d\tau \sum_{\alpha=1}^{N}
b^{\dagger}_{\alpha}(\tau) \partial_{\tau} b_{\alpha}(\tau)
+ {{1}\over{J}} \int^{\beta}_{0} d\tau \sum_{i=1}^{K} F_i^\dagger F_i \\
\nonumber  &&+  {{1}\over{N}} \int^{\beta}_{0} \!\!\!\! \int^{\beta}_{0} 
d\tau d\tau' \sum_{i\alpha}
F_i(\tau) b^{\dagger}_{\alpha}(\tau) G_{0}(\tau-\tau')
F_i^\dagger(\tau') b_{\alpha}(\tau')\\
&&+\int^{\beta}_{0} d\tau\, i\lambda(\tau)
 \left( \sum_{\alpha} b^{\dagger}_{\alpha}(\tau)b_{\alpha}(\tau) - p_0 N\right)
\label{action}
\end{eqnarray}
In this expression, 
$G_0(i\omega_n)\equiv\sum_{\vec{p}} 1/(i\omega_n-\epsilon_{\vec{p}})$
is the on-site Green's function associated with the conduction electron
bath. The quartic term in Eq.(\ref{action}) can be decoupled formally using 
two bi-local fields $Q(\tau,\tau')$ and $\overline{Q}(\tau,\tau')$ 
conjugate to 
$\sum_{\alpha}b^{\dagger}_{\alpha}(\tau)b_{\alpha}(\tau')$ and 
$\sum_{i} F_i^\dagger(\tau) F_i(\tau')$ respectively. Integrating out all 
other fields, the action can be solved by a saddle-point method over $Q$,   
$\overline{Q}$ and $\lambda$ for $N\rightarrow\infty$, which 
leads to coupled equations for the 
Schwinger boson and auxiliary field Green's functions
$G_b(\tau)\equiv -<Tb(\tau)b^\dagger(0)>$, 
$G_F(\tau)\equiv <TF(\tau)F^\dagger(0)>$
and for the Lagrange multiplier field (the latter is static 
and purely imaginary at the
saddle point: $i\lambda\equiv \bar{\lambda}$). 
These coupled equations read :
\begin{equation}
\Sigma_b(\tau)=\gamma G_0(\tau) G_F(\tau)\,\,\,,\,\,\,
\Sigma_F(\tau)= G_0(\tau) G_b(\tau)
\label{sp}
\end{equation}
where the self-energies $\Sigma_b$ and $\Sigma_F$ are defined by:
\begin{eqnarray}
\nonumber G_b^{-1}(i\nu_n)&=& i\nu_n+\bar{\lambda}-\Sigma_b(i\nu_n)\\
G_F^{-1}(i\omega_n) &=& {{1}\over{J}}-\Sigma_F(i\omega_n)
\label{defsigma}
\end{eqnarray}
In these expressions $\omega_n=(2n+1)\pi/\beta$ and $\nu_n=2n\pi/\beta$
denote fermionic and bosonic Matsubara frequencies.
Finally, $\bar{\lambda}$ is determined by the constraint:
\begin{equation}
G_b(\tau=0^-)\equiv \sum_n G_b (i\nu_n) e^{i\nu_n 0^+}\,=-\,p_0
\label{eqlambda}
\end{equation}

\begin{figure}[]
\epsfig{file=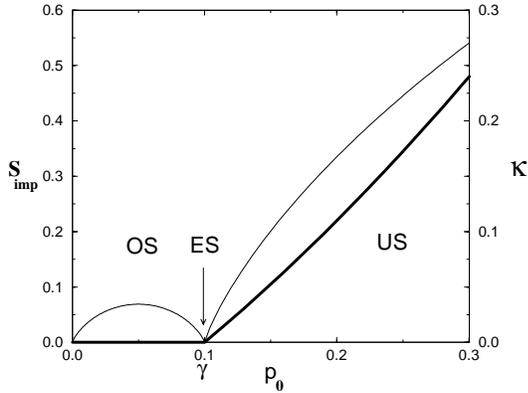,width=7cm,angle=0} 
\caption[0]{\narrowtext Residual entropy (thin line) 
 and Curie constant (bold line) {\it vs.} 
the ``size'' of the spin $p_0$, for $\gamma=0.1$ 
(analytical expressions given in the text).}
\label{entrocurie}
\end{figure}

We have studied these equations both numerically, and analytically in the 
limit of low-temperature and low-energy. The three Kondo regimes are best 
illustrated by Fig.\ref{entrocurie}, which displays the 
zero-temperature limit of the Curie constant 
$\kappa=\lim_{T\rightarrow 0} T\chi_{imp}(T)$ and of the impurity entropy 
$S_{imp}=\lim_{T\rightarrow 0} \lim_{V\rightarrow\infty}
[S(T)-S_{bulk}(T)]$, as a function of the ``size of the spin'' $p_0$. 
$\kappa$ vanishes in the \OS and \ES regimes ($p_0\leq\gamma$), but reaches 
a finite value $\kappa=(p_0-\gamma)(p_0-\gamma+1)$ in the \US regime 
($p_0>\gamma$), corresponding to the Curie constant of a residual 
spin of size $P_{sc}=P-K=N(p_0-\gamma)$ in the large-N limit.
Accordingly, the correlation function 
of the impurity spin does not vanish at long times in the \US regime: 
$<S(0)S(\tau)>\sim \hbox{const.}$, while 
$<S(0)S(\tau)>\sim 1/\tau^{2/(1+\gamma)}$ in the \OS regime. 
The latter is the behaviour expected from the conformal field theory (CFT)
 \cite{AL}
and Bethe Ansatz analysis \cite{Natan} 
of the non-Fermi liquid \OS fixed-point.   
The residual entropy at $T=0$ also takes the value expected from the 
degeneracy associated with a free spin of size $N(p_0-\gamma)$ in the 
\US regime, namely: 
$S_{imp}/N=(p_0-\gamma+1)\ln{(p_0-\gamma+1)} - (p_0-\gamma)\ln{(p_0-\gamma)}$,
 and vanishes at the \ES point.
In the \OS regime, its value is a universal function of $p_0$ and 
$\gamma$ associated 
with the non-Fermi liquid fixed point, which will be calculated explicitly 
below (Eq.\ref{Entropyover}). 

\begin{figure}[]
\epsfig{file=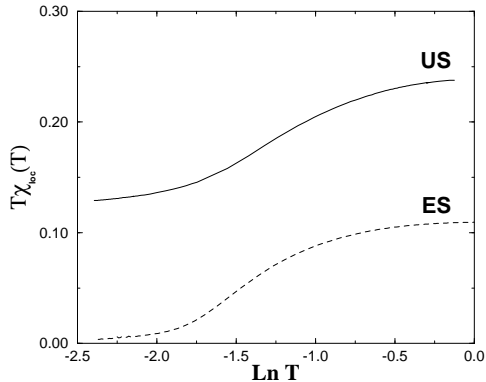,width=7cm} 
\caption[0]{\narrowtext Numerically calculated {\it local} Curie constant 
$T\chi_{loc}(T)$ {\it vs.} $\ln T$ with $\gamma=0.1$, for   
$p_0=0.1$ (\ES case) and $p_0=0.2$ (\US case)}
\label{kiloc}
\end{figure}

The cross-over between the high-temperature and low-temperature regimes, 
associated with the Kondo effect, 
is illustrated on Fig.\ref{kiloc} which displays the ``local Curie constant'' 
$T\chi_{loc}(T)$ vs $\ln T$
 in the \US and \ES regime (with $\chi_{loc}=\int_0^\beta <S(0)S(\tau)> 
d\tau $).

We now turn to a more detailed analysis of the low-energy, low-temperature 
behaviour of the above equations in the three regimes. 
In the \OS case $\gamma>p_0$, we first consider the zero-temperature limit, 
and perform a long-time asymptotic analysis
similar to that made in Ref.\cite{MH}. 
The crucial point in this regime is that the constant terms in the 
r.h.s of Eq.(\ref{defsigma}) vanish as $ T\rightarrow 0$, namely:
$\bar{\lambda}-\Sigma_b(i0^+)\rightarrow 0, 1/J-\Sigma_F(i0^+)\rightarrow 0$.
 As a result, a power-law decay 
$G_b(\tau)\sim A_b/\tau^{2\Delta_b}\,,\,
G_F(\tau)\sim A_F/\tau^{2\Delta_F}$ is found for the Green's 
functions in the limit $T_{K}^{-1}\ll\tau\ll\beta\rightarrow\infty$ (where 
$T_{K}$ is the Kondo temperature). 
The exponents are given by:
$2\Delta_b = 1/(1+\gamma)\,,\, 2\Delta_F = \gamma/(1+\gamma)$, so that 
the local spin-spin correlation decays as 
$\langle S(0)S(\tau)\rangle \propto G_b(\tau)G_b(-\tau)\sim
 1/\tau^{2/(1+\gamma)}$.  
The spectral density of the Schwinger boson has the following 
low-frequency behaviour at zero temperature (with 
$C$ a positive constant):
\begin{equation}
\rho_{b}(\omega\rightarrow 0^{\pm})\sim C
{{\sin{\theta_{\pm}}}\over{(\pm\omega)^{1-2\Delta_b}}}
\label{specb}
\end{equation}
It is important to observe that the same power-law behaviour is found 
for $\omega>0$ and $\omega<0$, but with {\it asymmetric prefactors} 
parametrised 
by the phases $\theta_{\pm}$. The values of these phases can be found using 
a Luttinger-Ward argument \cite{long1,long2}, leading to:
$\sin{\theta_+}=\sin{ {{\pi}\over{1+\gamma}}(\gamma-p_0)}\,,\,
\sin{\theta_-}=-\sin{ {{\pi}\over{1+\gamma}}p_0}$. From these expressions, 
we see that Eq.(\ref{specb}) obeys the positivity 
property appropriate for a bosonic spectral density:  
$\mbox{sign}(\omega)\rho_b(\omega)>0$
{\it only as long as} $p_0<\gamma$ ($P<K$). The violation 
of this property signals the transition to the \US regime, in 
which the solutions just described are no longer valid.

Contact can be made, at least qualitatively, 
with the usual NCA equations (enforcing $P=1$ strictly), 
and with Ref.\cite{CR}, by taking 
the limit $p_0\rightarrow 0$, {\it i.e} dealing with an impurity such that  
$P\ll N$. In that case, we observe that 
$\theta_-\rightarrow 0$, and the $T=0$ spectral density vanishes for 
negative frequencies. There is no such threshold for non-zero values 
of $p_0$ however, but the spectral density does become increasingly asymmetric 
as $p_0$ gets smaller. 

In order to calculate the low-temperature behaviour of thermodynamic 
quantities, the above $T=0$ form of the spectral densities is 
insufficient however. The limit $\omega,T\rightarrow 0$ must be 
taken while keeping $\tilde{\omega}\equiv \omega/T$
 finite \cite{Subir}. We have succeeded 
\cite{long1,long2} in 
calculating analytically the spectral functions in this limit, which 
takes a {\it scaling form}:
$\rho_{b,F}(\omega) = A_{b,F} T^{2\Delta_{b,F}-1}\,
\phi_{b,F}({\tilde\omega})$.
The scaling function $\phi_{b}$ (found from either a direct solution 
of the integral equations above, or from conformal field theory arguments)
reads:
\begin{equation}
\phi_b(\tilde{\omega};p_0)=
{(2\pi)^{2\Delta_b} \over{2\pi^2}}\sinh{{\tilde{\omega}} \over{2}}\,
{{|\Gamma(\Delta_b+i{{\tilde{\omega}-\tilde{\omega}_0}\over{2\pi}})|^2}
\over{\Gamma(2\Delta_b)}}
\label{phib}
\end{equation}
in which $\tilde{\omega}_0\equiv\ln{|\sin{\theta_+}/\sin{\theta_-}|}$. 
$\phi_F$ has a similar expression, with $\Delta_F$ replacing 
$\Delta_B$ and $\cosh{\tilde{\omega}/2}$ replacing 
$\sinh{\tilde{\omega}/2}$.
Using Eq. (\ref{phib}) and the expression of the free energy 
$ F_{imp}/N =  - p_0 \bar{\lambda}- \,T\sum_n \mbox{ln} G_b(i\nu_n) +
\gamma\, T\sum_n \mbox{ln} G_F(i\omega_n)
 - \,T\sum_n \Sigma_{b}(i\nu_{n})G_{b}(i\nu_{n}) $
we have obtained \cite{long1,long2} the zero-temperature 
limit of the entropy, which is a universal number associated with the 
non-Fermi liquid fixed point \cite{ALent}. We find:
\begin{equation}
{{1}\over{N}}S_{imp}= {1+\gamma\over \pi} 
\left[ f_{\gamma} (1+p_{0})
- f_{\gamma} (1) - 
f_{\gamma} (p_0) \right]
\label{Entropyover}
\end{equation}
with $f_{\gamma}(x) \equiv \int_0^{\pi x\over 1+\gamma} \ln\sin (u)\,d u$. 
We have also calculated \cite{long1,long2} 
$S_{imp}$ for arbitrary finite values 
of $N,K$ and $P$ by applying the conformal-field theory (CFT) methods of 
Ref.\cite{ALent,AL}. We find (for $K>P$)
$S_{imp}\,=\, \mbox{ln}\,\prod_{n=1}^{P}
\sin {{\pi (N+n-1)}\over{N+K}}/\sin {{\pi n}\over{N+K}}$, 
which can also be established using Bethe Ansatz methods \cite{Natan}.
The large-N limit of this expression coincides with (\ref{Entropyover}), 
with corrections of order $1/N$ \cite{KC}. 
We have also computed the low-temperature behaviour of the specific heat 
ratio and impurity susceptibility in the overscreened regime, and found :
$C/T\sim\chi_{imp}\sim 1/T^{(\gamma-1)/(\gamma+1)}$ for $\gamma>1$, 
$C/T\sim\chi_{imp}\sim \mbox{ln}1/T$ for $\gamma=1$, 
$C/T\sim\chi_{imp}\sim \mbox{const.}$ for $\gamma<1$, also in agreement 
with the CFT and Bethe Ansatz results. Note that the last behaviour holds even 
though the fixed point is {\it not} a Fermi liquid (as evidenced by 
the fact that $S_{imp}\neq 0$).
 
In the \US regime, the crucial difference with the above analysis is 
that $1/J-\Sigma_F(i0^+)$ no longer vanishes at zero-temperature, 
but instead {\it diverges logarithmically}: 
$1/J-\Sigma_F(i0^+)\sim  {1/J_e} \ln T + J_2$ 
(while $\bar{\lambda}-\Sigma_b(i0^+)\sim \lambda_1 T$). 
This behaviour is well obeyed by the numerical solution 
of the above equations, and has a simple physical interpretation:
in the \US case, the residual 
spin is asymptotically free at low temperature, but is weakly coupled to 
the conduction electron gas at any finite temperature by a 
{\it ferromagnetic} Kondo coupling which vanishes logarithmically as 
temperature is reduced. Based on this observation, we can find the 
low-temperature scaling form of the spectral functions as: 
$\rho_b(\omega)=\delta(\omega/Z+\lambda_1 T)+\cdots$ and 
$\rho_F(\omega)=1/\ln^2T\phi_F(\omega/T)+\cdots$. The form of $\rho_b$ 
is characteristic of an (asymptotically) free local spin. The detailed 
expression of $\phi_F$ will be given elsewhere \cite{long2}.   

Finally, we briefly comment on the \ES case. 
The large-N limit preserves the local impurity properties 
expected from the full Kondo screening, namely $S_{imp}\rightarrow 0$ 
and $\chi_{loc}\rightarrow \mbox{const.}$ as $T\rightarrow 0$. 
The conduction electrons form a local Fermi-liquid which is however 
only weakly affected by the
screening of the impurity in the large-N limit. 
Indeed, the scattering phase shift obtained from 
Friedel's sum rule is $\delta/\pi=1/N$. 
Accordingly, the scattering $T$-matrix, obtained in 
our approach as the Fourier transform of $G_b(\tau)G_F(-\tau)$, 
is found to have 
vanishing imaginary part in the limit of $\omega,T\rightarrow 0$.  
This is a peculiarity of the 
specific spin representation that we have considered: considering rectangular  
Young tableaus with $P=K=\gamma N$ columns and $N/2$ lines would 
restore maximal unitary scattering $\delta=\pi/2$ for all $N$.
Whether a large-N limit exists for that case also is an interesting 
open problem, with potentially useful applications 
{\it e.g} to the Kondo lattice model.

During the course of this study, we learned of simultaneous work by
N.Andrei, A.Jerez and G.Zar{\'a}nd 
on the same model using the Bethe Ansatz method \cite{Natan}.
Our results and conclusions agree when a comparison is possible.  
We are most grateful to N.Andrei for numerous
discussions.
We also thank T.Giamarchi, P.Coleman, G.Kotliar, 
A.Sengupta, A.Ruckenstein and P.Zinn-Justin for very helpful suggestions.

\end{multicols}

\begin{references}
\bibitem{foot} In all three regimes, the strong-coupling fixed point 
involves $(N-1)K$
conduction electrons, which combine into an $SU(N)$ representation
corresponding to a Young tableau with $N-1$ lines and $K$ columns.
\end{references}
\end{document}